# Tunable Magnetic Properties in $Sr_2FeReO_6$ Double-Perovskite

Zhaoting Zhang[1,2,○], Hong Yan[1○], Zhen Huang[3]*, Xiao Chi[1,4], Changjian Li[5], Zhi Shiuh Lim[1], Shengwei Zeng[1], Kun Han[3], Ganesh Ji Omar[1], Kexin Jin[2]*, Ariando Ariando[1]*

[1]*Department of Physics, National University of Singapore, 117575, Singapore*

[2]*Shaanxi Key Laboratory of Condensed Matter Structures and Properties and MOE Key Laboratory of Materials Physics and Chemistry under Extraordinary Conditions, Northwestern Polytechnical University, Xi'an 710072, China*

[3]*Information Materials and Intelligent Sensing Laboratory of Anhui Province, Institutes of Physical Science and Information Technology, Anhui University, 230601, China*

[4]*Singapore Synchrotron Light Source, National University of Singapore, 117603, Singapore*

[5]*Department of Materials Science and Engineering, Southern University of Science and Technology, 518055, China*

*E-mails: ariando@nus.edu.sg; jinkx@nwpu.edu.cn; huangz@ahu.edu.cn
○These authors contributed equally.

**ABSTRACT:** Double-perovskite oxides have attracted recent attention due to their attractive functionalities and application potential. In this paper, we demonstrate the effect of dual controls, i.e., the deposition pressure of oxygen ($P_{O_2}$) and lattice mismatch ($\varepsilon$), on tuning magnetic properties in epitaxial double-perovskite $Sr_2FeReO_6$ films. In a nearly-lattice-matched $Sr_2FeReO_6/SrTiO_3$ film, the ferrimagnetic-to-paramagnetic phase transition occurs when $P_{O_2}$ is reduced to 30 mTorr, probably due to the formation of $Re^{4+}$ ions that replace the stoichiometric $Re^{5+}$ to cause disorders of B-site ions. On the other hand, a large compressive strain or tensile strain shifts this critical $P_{O_2}$ to below 1 mTorr or above 40 mTorr, respectively. The observations could be attributed to the modulation of B-site ordering by epitaxial strain through affecting elemental valence. Our results provide a feasible way to expand the functional tunability of magnetic double-perovskite oxides that hold great promise for spintronic devices.

The large variety of possible cation combinations has made double-perovskite oxides a material platform for designing new and enhanced physical properties for future applications.[1-10] Examples are colossal magnetoresistance effect for disk read-and-write heads,[7] half-metallicity for single-spin electron measurement,[1,8] and multiferroicity for electrically-driven memory devices with low energy consumption.[2-4] Among various double-perovskite oxides, the Re-based double-perovskites are appealing due to their high Curie temperature,[9] large spin polarization[11] and strong spin-orbit coupling,[12] potential for room-temperature magnetoresistive and spintronic devices.[10] One of the most prominent double perovskite oxides is $Sr_2FeReO_6$ (SFRO), which exhibits high Curie temperature (~400 K) along with a high degree of spin-polarization having half-metallic behavior.[10] In SFRO, each $Fe^{3+}$ ion ($3d^5$, $S = 5/2$) is surrounded by six $Re^{5+}$ ($5d^2$, $S = 1$) ions or vice versa and couple to each other antiferromagnetically. The SFRO compounds with heterovalent transition metals occupying B exhibit rich electronic structures and complex magnetic structures owing to the strong interplays between strongly localized 3d electrons and more delocalized 5d electrons with strong spin-orbit coupling.

Generally, the physical properties of double-perovskite oxides can be systemically tuned by selecting different A- and B-site cations, which determine the electronic band structure of materials.[13-15] Both $Ba_2CdReO_6$ and $Ba_2FeMoO_6$ are predicted to be a magnetic topological quantum material.[20] However, due to the difference in B-site cations, $Ba_2CdReO_6$ is an intrinsic ferromagnetic topological nodal-ring semi-half-metal with a pair of Weyl points and one nodal ring in the first Brillouin zone, while $Ba_2FeMoO_6$ belongs to a ferrimagnetic topological half-metal, having four pairs of Weyl points and two fully spin-polarized nodal rings.

The design of two-dimensional multiferroic materials with intrinsic magnetoelectric coupling based on double-perovskites oxides has been proposed, in which two coupling mechanisms between polarization and magnetization have been found to enable the reversal of the in-plane magnetization by ferroelectric switching.[21] Meanwhile, the synthesis parameters (such as the deposition pressure of oxygen $P_{O_2}$ and growth temperature) also play important roles in property tuning.[16-18] For the $Sr_2FeReO_6/SrTiO_3$ (SFRO/STO) films, a recent study has shown that the low (high) $P_{O_2}$ induces additional Re (Fe) ions and eradicates the B-site cation ordering, leaving only a small growth window to obtain the stoichiometric, B-site-ordered and ferrimagnetic conducting SFRO films.[19] On the other hand, the lattice-mismatch-induced epitaxial strain can modify the lattice structure to provide the functional tunability in complex oxide heterostructures.[22-25] Using density functional theory together with Monte Carlo simulations, it has been demonstrated that epitaxial strain, both compressive and tensile, attenuates the spin frustration of double perovskite $Sr_2FeOsO_6$ to significantly enhance the critical temperature to 310 K[26] and a tetragonal-to-monoclinic structural transition concomitant with an antiferromagnetic-to-ferrimagnetic transition under tensile strain. Meanwhile, a strain-induced change of magnetic properties has been revealed in $Sr_2CoIrO_6$ films. By reversing the applied strain direction from tensile to compressive, the easy axis changes abruptly from in-plane to out-of-plane orientation.[27] In this report, we focus on the strain response of double-perovskite SFRO films, of which the strain is mediated by choosing different substrates including (001)-oriented $LaAlO_3$ (LAO, with a lattice constant of 3.792 Å), $(La_{0.3}Sr_{0.7})(Al_{0.65}Ta_{0.35})O_3$ (LSAT, 3.868 Å), $SrTiO_3$ (STO, 3.905 Å), and $KTaO_3$ (KTO, 3.988 Å).[28]

The lattice structure of ordered double perovskite SFRO can be understood by alternatingly stacking corner-sharing $FeO_6$ and $ReO_6$ octahedra, as shown in **Figure 1**(a). The electronic configurations of Re and Fe are $[Xe]4f^{14}5d^56s^2$ and $[Ar]3d^64s^2$, respectively. In the SFRO films, the Fe cations prefer to maintain the most stable +3 state, while the Re ions are expected to stay in +5 state to keep the charge neutrality in stoichiometric SFRO films. Meanwhile, the +3 Fe cations will show a fully occupied spin-up band[+] ($3d^5$: $t_{2g}^3\uparrow$ $e_g^2\uparrow$),[29] accompanied by the formation of an empty spin-down band around 1 eV above the spin-up one.[30] Given that the $Re^{5+}$ ($5d^2$) cations are antiferromagnetically coupled with neighbor $Fe^{3+}$ ($3d^5$) ones,[31] the Fermi energy lies in the hybridized spin-down channel consisting of Fe 3d and Re 5d orbitals to provide the fully spin-polarized conductivity, as shown in Figure 1(b). Hence, the SFRO is expected to show the half metallicity. Also, because both $Fe^{3+}$ ($3d^5$) and $Re^{5+}$ ($5d^2$) cations are in the high-spin state, the ideal saturated magnetic moment ($M_S$) of ferrimagnetic SFRO is 3 $\mu_B$ per formula unit cell ($\mu_B$/f.u.). However, the experimental observed $M_S$ is always much lower than the expected value, due to the formation of antisite defects that has been widely observed in various double-perovskites.[32-34] In Figure 1(c), the field-dependent isothermal magnetizations, $M$(H), are measured at 300 K for SFRO films grown on different substrates with a fixed $P_{O_2}$ = 30 mTorr. Compared to the bulk SFRO with lattice constants of $a_p$ = 7.864 Å, $c_p$ = 7.901 Å,[35] the LAO, LSAT, STO and KTO substrates provide the nominal in-plane lattice mismatch of −3.6%, −1.6%, −0.7% and +1.4% to the epitaxial SFRO layer, respectively. Among those SFRO thin films, the lattice-matched SFRO/STO films with optimized $P_{O_2}$ (around 30 mTorr) are expected to show the stoichiometric composition with B-site Fe/Re ordering and ferrimagnetic state.[21] This is consistent with our results on SFRO/STO samples, where the B-site Fe/Re ordering is confirmed by the (111) diffraction peak recorded by the off-axis X-ray diffraction (XRD) (Figure

S1b), accompanied by the clear $M(H)$ hysteresis with a saturation magnetization of 0.8 $\mu_B$/f.u. at 300 K in Figure 1c. When comparing to the lattice matched SFRO/STO film, both the compressive and tensile strain suppress the ferromagnetic interaction and reduce $M_S$ to 0.5 and 0.2 $\mu_B$/f.u. with a fixed $P_{O_2}$ =30 mTorr, respectively. Those results indicate that the epitaxial strain may act as another parameter, as compared to the oxygen pressure, to control the magnetic property of double-perovskite SFRO layers.

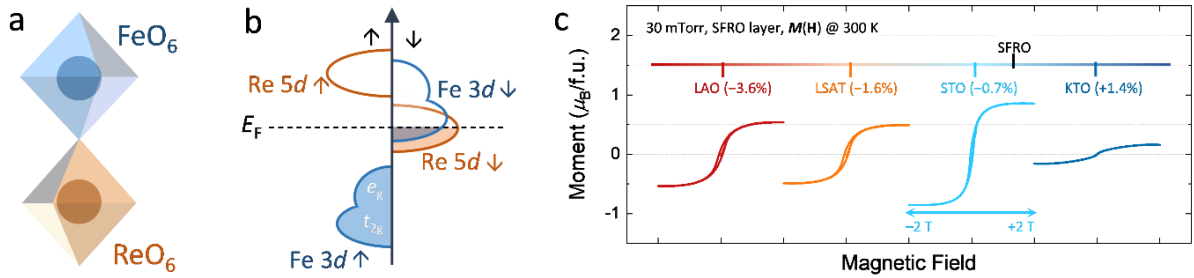

**Figure 1.** a. The arrangement of FeO$_6$ and ReO$_6$ octahedra in the crystal structure of double perovskites SFRO. B. Sketch of the band structure for the illustration of the ferrimagnetic state in the double perovskites SFRO. C. Isothermal magnetization of SFRO films grown at different substrates with applied in-plane external magnetic field at 300 K.

In **Figure 2**, we demonstrate the dual controls of the oxygen pressure and epitaxial strain on the magnetic SFRO layers. Figure 2(a) compares $M(H)$ curves of samples deposited at wide-ranged $P_{O_2}$ for LSAT and STO substrates at 10 K. The $M(H)$ curves for the SFRO films grown at high $P_{O_2}$ ($\geqslant$ 30 mTorr) on STO substrates show clear hysteresis behavior, while give rise to paramagnetic ground state at low $P_{O_2}$ ($\leqslant$ 20 mTorr). This is nearly consistent with previous reports.[21] The maximal $M_S$ is 1.8 $\mu_B$/f.u. at 10 K for the sample grown at 40 mTorr. This identifies the antiferromagnetic coupling between Fe$^{3+}$ and Re$^{5+}$ on the B sites, as the ferrimagnetic state should produce $M_S$ as the difference of the two moments, which is 3 $\mu_B$/f.u. Such a difference between the ideal and observed values is principally the consequence of antisite disorders at B sites.[36-38] In terms of the empirical relationship between $M_S$ and the amount of antisite disorder, the observed optimal $M_S$ implies a defect concentration of about 10% in SFRO films on STO substrates with the $P_{O_2}$ of 40 mTorr.[39] For LSAT substrate, $M(H)$ hysteresis loops is observed at all $P_{O_2}$. For low $P_{O_2}$ sample, the LSAT-based films with the compressive strain still maintain the ferrimagnetic ground state. This suggests compressive strain may promote the appearance of B-site ordered SFRO perovskites. Temperature dependence of sheet resistance ($R_s$) for SFRO films is shown in Figure 2b. The $R_s$(T) curves for low $P_{O_2}$ samples on both substrates present a very slow growth with decrease of temperature, indicating that the resistivity of samples is dominated by charge localization, as widely observed for the polycrystalline double perovskites.[40] According to the difference of magnetic property of low $P_{O_2}$ films on STO and LSAT substrates, obviously magnetic interaction seems to be weakly coupled with mobile electrons. With increasing $P_{O_2}$, a semiconducting tendency is developed and eventually evolved into an insulating behavior for the films grown on both substrates. The observations suggest that the sample conduction is not directly correlated with the magnetic property. The change of resistance at room-temperature is about four orders of magnitude. The films on STO substrates show slightly lower $R_s$ compared to that on LSAT substrates, indicating the decrease of metallic grain boundaries.

To further investigate the effect of epitaxial strain on magnetism, we studied magnetic behaviors for the SFRO films on other substrates and summarized in Figure 2c. The photographs of samples grown on LSAT substrates are shown in the right panel. The pure LSAT substrate is transparent and colorless. After being deposited with SFRO films, its color gradually darkens with decreasing $P_{O_2}$. For low $P_{O_2}$, SFRO films on LSAT and LAO substrates exhibit ferrimagnetic ground state, while that on STO and KTO substrates present paramagnetic behavior. With increasing $P_{O_2}$, the $M_S$ finally reaches peak value, before presenting a slight decrease. And the factor of $P_{O_2}$ gives rise to more significant response in weakly compressed LSAT and STO based films, compared to largely compressed LAO and tensile KTO substrates. The modulation ranges for LSAT and STO substrates are about 1.69 and 2.03 $\mu_B$/f.u., while that of LAO and KTO are about 0.91 and 0.88 $\mu_B$/f.u., respectively. This result can be obviously attributed to strain effect. It has been demonstrated that the $P_{O_2}$ has a significant impact on B-site cation ordering, known to be mainly dominated by the difference of the ionic radii of two cations in double-perovskite systems.[21] It has been observed that for STO-based films the valence of Fe ions remains as 3+ while the valence of Re ions changes from 4+ in low $P_{O_2}$ to 6+ in high $P_{O_2}$ to satisfy charge neutrality. Compressive strain can suppress the formation of $Re^{4+}$ at low $P_{O_2}$, because $Re^{4+}$ ions have significantly larger ionic radii of 77 pm than that of $Re^{5+}$ and $Re^{6+}$ ions of 72 and 69 pm respectively. However, for KTO-based films, tensile films contribute more to the appearance of $Re^{5+}$ and $Re^{6+}$ ions having ionic radii of 72 and 69 pm. Our observation supports this explanation that the magnetization of KTO-based films reaches maximum at about 80 mTorr, while the films on other substrates get maximum at 40 mTorr. In LAO-based films, despite the large compressive stress, the huge lattice mismatch may destroy the ordered arrangement of cations at the B site, resulting in a reduction in magnetization.

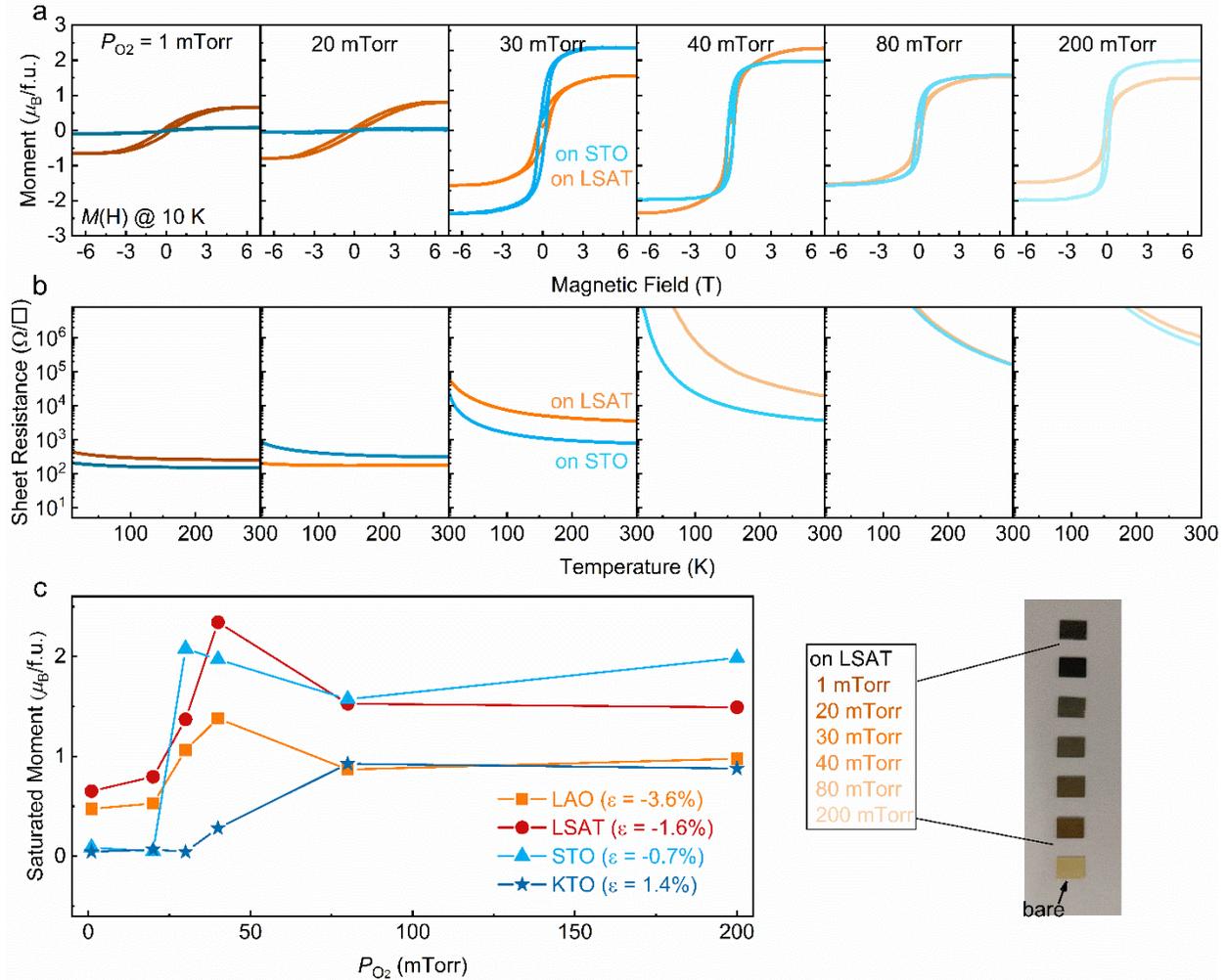

**Figure 2.** a. Isothermal magnetization of SFRO films grown at various oxygen pressure on LSAT and STO substrates with applied in-plane external magnetic field at 10 K. b. Temperature dependence of sheet resistance for SFRO films grown at various oxygen pressure on LSAT and STO substrates from 10 K to 300 K. c. Saturated magnetization of SFRO films grown at various oxygen pressure on LAO, LSAT, STO and KTO substrates with in-plane magnetic field at 10 K. The right of panel shows photo of samples on LSAT substrate with pristine LSAT substrate as comparison.

To gain insights into structural and chemical distributions on the atomic scale, scanning transmission electron microscopy (STEM) and Fast Fourier Transform (FFT) analyses were carried out on the SFRO films on LSAT grown at 20 mTorr (**Figure 3**). The FFT diffraction pattern gives two set of dot matrixes, demonstrating two double perovskite structure. The high-angle annular dark field (HAADF) image can be classified as two kinds of different areas. In one area B-site atoms seem to periodically dim (Fe atoms) and brighten (Re atoms), called highly ordered area marked by orange square. The corresponding line intensity profile shows alternately arranged high and low peaks, as shown in Figure 3c. High peaks demonstrate Re atom prevails, while low peaks are dominated by Fe atoms. In the other area undifferentiated brightness changes are presented. The line profile also proves this result (Figure 3d). This hints Fe and Re atoms at B-site atoms are not alternately arranged. This result demonstrates the keep of B-site ordering at compressive-strained SFRO/LSAT films.

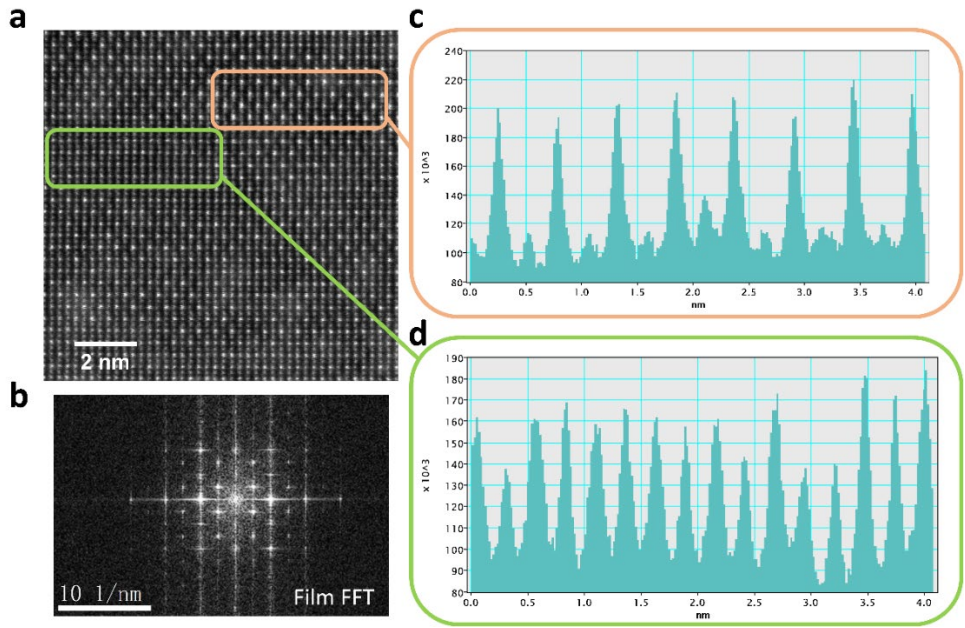

**Figure 3.** a. HAADF-STEM images for $Sr_2FeReO_6$ film on LSAT grown at 775 °C under 20 mTorr. Highly and less ordered areas are marked by orange and green squares, respectively. b. Corresponding FFT diffraction pattern for SFRO film. c and d. Line intensity profile extracted across the B site atoms in highly and less ordered areas indicated with orange and green squares.

The XRD of samples grown at 1 mTorr is shown in **Figure 4**. The (004) peak positions of SFRO film deposited on KTO, STO, LSAT, and LAO substrates are 46.13°, 45.14°, 44.74°, and 45.58°, respectively. According to the position of the peak, we can calculate the out-of-plane lattice constant of the films, which are 3.929, 4.014, 4.048, and 3.976Å, respectively. The out-of-plane lattice constant gradually increases as the strain changes from tensile strain to compressive strain, except for LAO case. Obviously, the SFRO films on KTO, STO, and LSAT can still maintain the coherent growth, while the film on LAO shows structural relaxation tendency to its bulk value. Therefore, the volumes of SFRO unit cell on STO and LSAT will be 31.349 Å$^3$, and 31.315 Å$^3$. Compared with STO case, the decrease of SFRO cell volume will suppress the change of Re$^{5+}$ to Re$^{4+}$, resulting in the occurrence of ferromagnetism of SFRO film on LSAT substrate. Figure 4b shows XRD off-axis scans at ψ = 47.5° near the SFRO (111) peak, which is the indicator of B-site cation ordering. This peak is absent on the STO substrate, while it exists on the other cases. Its existence for LSAT case confirms the Fe/Re ordering and accounts for the appearance of magnetic hysteresis while absent for SFRO film on STO substrate.

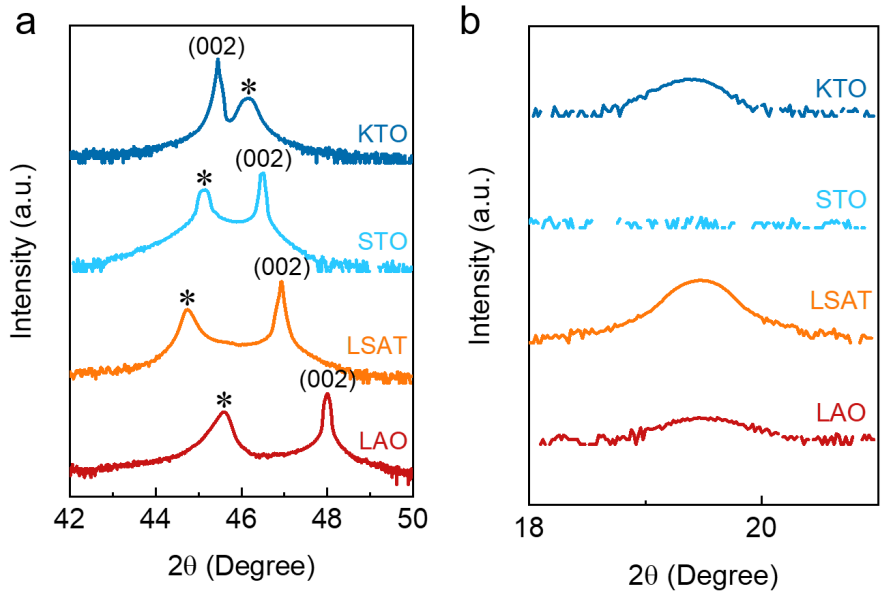

**Figure 4.** a. XRD θ-2θ scans for SFRO films on different substrates grown at 1 mTorr. The (004) peak of SFRO films are marked by star. b. Corresponding off-axis XRD θ-2θ scans near the SFRO (111) peak.

To explain observed phenomenon more clearly, different structures and corresponding energy level diagram are presented in **Figure 5**. It's noted that $P_{O_2}$ mainly controls the relative ratio between Fe and Re cations, while barely affecting the Sr and O stoichiometry. In stoichiometric SFRO films, intervening oxygen connects every Fe and Re ion pair, thus forming alternating FeO$_6$ and ReO$_6$ octahedra. Fe$^{3+}$ and Re$^{5+}$ ions arrange alternately on the B-sites of double perovskite with respective spins coupling antiferromagnetically. As shown in Figure 5a, the occupied up-spin channel is mainly occupied by Fe 3d electrons hybridized with O 2p states. The Re t$_{2g}$ and e$_g$ up-spin channels are above the Fermi level. By comparison, the down-spin channel around the Fermi level is mainly composed of both the Re 5d $t_{2g}$ and Fe 3d $t_{2g}$ electrons, which are strongly hybridized with O 2p states. Such a half-metallic nature gives rise to fully spin-polarized charge carriers in the ground state. The epitaxial strain has a small effect on the films in this region (30mTorr ≤ $P_{O_2}$ ≤ 40 mTorr) except for the decreased saturation magnetic moment. At low $P_{O_2}$ region (≤ 20 mTorr), a Re-rich paramagnetic phase has been demonstrated for STO based SFRO films.[19] As has been noted, the valence of Re ions changes to 4+ in Re-rich phases to satisfy charge neutrality while the valence of Fe ions remains as 3+ in all samples as shown in Figure 5b. However, ionic radius of $Re^{4+}$ ions is 77 pm, which is very close to the ionic radius of Fe ions (78.5 pm). This will lead to the disappearance of B-site cation ordering and then the absence of long-range ordering in Re-rich films as observed in Figure 2. However, in LSAT based films, compressive strain suppresses this process. The change from $Re^{5+}$ to $Re^{4+}$ ions is slow down as the larger ionic radius of $Re^{4+}$ than that of $Re^{5+}$ (72 pm). Therefore, a certain number of long-range ordering is preserved and presented as weak magnetic hysteresis loop. On the contrary, in KTO based films, epitaxial tensile stress dominates. It will benefit the formation of $Re^{4+}$ ions. This is consistent with our observation in Figure 2 that for KTO based films they have larger critical $P_{O_2}$ where there is the maximum saturated moment. At high $P_{O_2}$ region (≥ 40 mTorr), a Fe-rich ordered phase has been demonstrated. To satisfy charge neutrality, valence of Re ions changes to 6+ even 7+ in Fe-rich phases. Since ionic radii of $Re^{6+}$ (69 pm) or $Re^{7+}$ (67 pm) ions are much smaller than the ionic radii of Fe ions (78.5 pm), ordered structure remains. Then ferrimagnetic ground state doesn't change. But extra Fe forms $Fe^{3+}$-$Fe^{3+}$ bonding, which is antiferromagnetic coupled. The appearance of antiferromagnetic phases will inevitably lead to the reduction of saturation magnetic moment, which is consistent with the observation in Figure 2. Meanwhile, the addition of Re valence also results in the reduction of conduction band electron as shown in Figure 5(c) and then the increase of resistivity. This result also coincides with the increase of resistance with oxygen pressure.

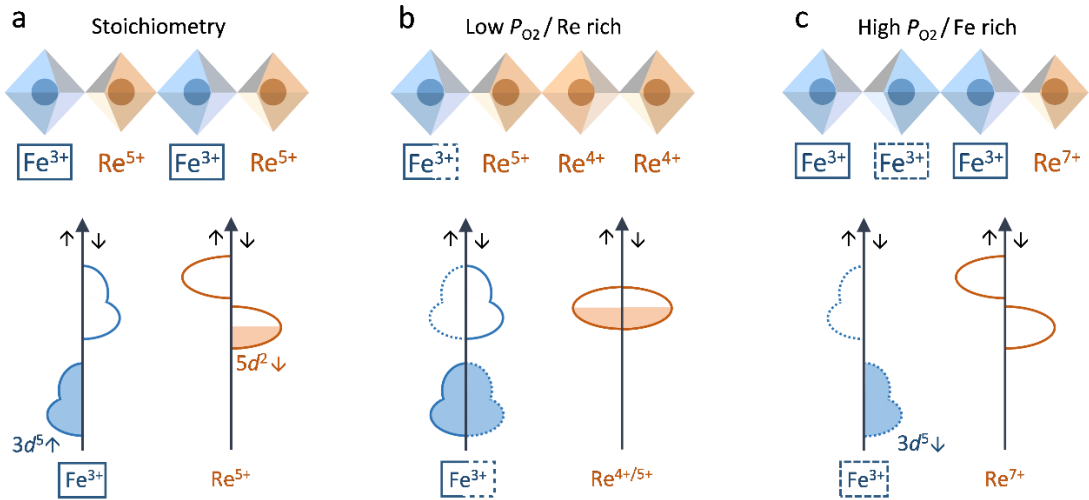

**Figure 5.** a, b, and c The arrangement of FeO$_6$ and ReO$_6$ octahedra and corresponding band structure in double perovskites SFRO films with the stoichiometric, Re-rich, and Fe-rich crystal structure, respectively. In b and c, the Fe$^{3+}$ ions bordered by dotted lines refer to those with antiparallel moment alignment with the solid-line bordered ones.

In summary, we have the dual control of magnetic properties in SFRO double perovskites by growth parameter $P_{O_2}$ and lattice mismatch. Importantly, ferrimagnetic-to-paramagnetic phase transition of SFRO films could be modulated effectively by mismatch strains imposed by the substrate. With increasing tensile stain, the critical $P_{O_2}$ of the transition moves to high $P_{O_2}$, but moves to low $P_{O_2}$ under compressive strain. Our results indicate that the ordering of B-site cations in double perovskite SFRO films could be changed by controlling growth parameter $P_{O_2}$ and epitaxial strain. This new discovery may advance our understanding of the relationship among strain engineering, synthesis conditions and physical properties in double perovskites and explore their potential applications in spintronic devices.

**Supporting Information**

The Supporting Information is available free of charge at https://xxx

Materials and methods; sample preparation; transport and magnetic measurement; XRD and STEM imaging; X-ray diffraction data; magnetoresistance; temperature dependent magnetization; and normalized resistance of SFRO thin film (PDF)

**Author Contributions**

A.A. conceived and led the project. Z.Z. supported by H.Y., Z.H., X.C., Z.S.L., S.Z., K.H., and G.J.O. prepared the samples and performed electrical, magnetic, and structural characterizations. C.L. performed STEM experiments and analysis. Z.H. and K.J. supported the project supervision. All authors discussed the results and modified the manuscript.

**Notes**

The authors declare no competing financial interest.

**ACKNOWLEDGMENTS**

This research is supported by the Agency for Science, Technology and Research (A*STAR) under its Advanced Manufacturing and Engineering (AME) Individual Research Grants (IRG) (A1983c0034 & A2083c0054). Z.S.L. and A. A. acknowledge the National Research Foundation (NRF) of Singapore under its NRF-ISF joint program (Grant No. NRF2020-NRFISF004-3518) for the financial support. Z.Z. and K.J. acknowledge the financial support from the National Natural Science Foundation of China (Nos. 51572222, 11604265), Key Research Project of the Natural Science Foundation of Shaanxi Province, China (Grant No. 2021JZ-08), and the scholarship from China Scholarship Council (CSC) under the Grant No. CSC201906290191.

TOC Graphic

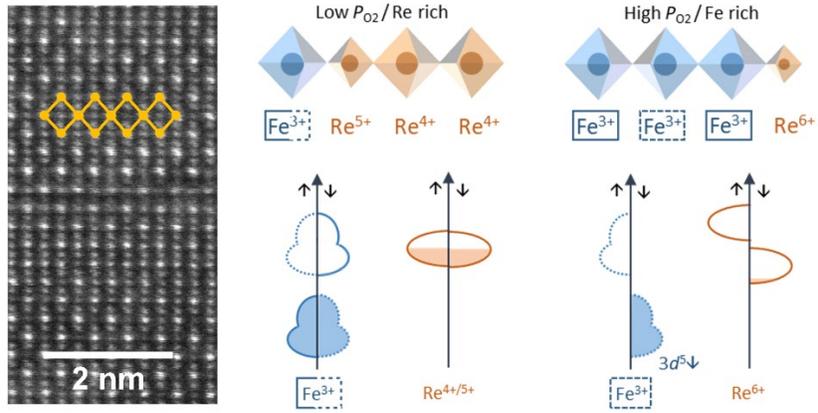

**Supporting Information**

# Tunable Magnetic Properties in $Sr_2FeReO_6$ Double-Perovskite

Zhaoting Zhang[1,2○], Hong Yan[1○], Zhen Huang[3]*, Xiao Chi[1,4], Changjian Li[5], Zhi Shiuh Lim[1], Shengwei Zeng[1], Kun Han[3], Ganesh Ji Omar[1], Kexin Jin[2]*, Ariando Ariando[1]*

*[1]Department of Physics, National University of Singapore, 117575, Singapore*

*[2]Shaanxi Key Laboratory of Condensed Matter Structures and Properties and MOE Key Laboratory of Materials Physics and Chemistry under Extraordinary Conditions, Northwestern Polytechnical University, Xi'an 710072, China*

*[3]Information Materials and Intelligent Sensing Laboratory of Anhui Province, Institutes of Physical Science and Information Technology, Anhui University, 230601, China*

*[4]Singapore Synchrotron Light Source, National University of Singapore, 117603, Singapore*

*[5]Department of Materials Science and Engineering, Southern University of Science and Technology, 518055, China*

*E-mails: ariando@nus.edu.sg; jinkx@nwpu.edu.cn; huangz@ahu.edu.cn

○These authors contributed equally.

Keywords: oxide heterostructures, double perovskites, epitaxial strain, magnetism

**Materials and Methods**

***Sample Preparation:*** The SFRO films were epitaxially deposited on LAO, LAST, STO, and KTO single crystalline substrates by pulsed laser deposition (PLD) technique. A commercial sintered polycrystalline ceramic bulk of Sr$_2$FeReO$_6$ was used as the target. Before deposition, PLD chamber was firstly evacuated to a high vacuum condition (~10$^{-6}$ mTorr), and then high purity oxygen was introduced into the chamber to reach desired $P_{O_2}$. The $P_{O_2}$ was varied from 1 to 200 mTorr to obtain SFRO thin films with different oxygen stoichiometry. During deposition, laser energy density was kept at ~1.8 J/cm$^2$, temperature at 775 °C and a pulse repetition rate of 5 Hz with a KrF excimer laser of 248 nm in wavelength. After deposition, the SFRO films were slowly cooled down to room temperature at the rate of 10°C/min. The PLD chamber was always kept at desired $P_{O_2}$ during the whole growth process. All the films used in this work were about 80 nm in thickness.

***Transport and magnetic measurement:*** All the magnetic measurements were performed on Quantum Design Superconducting Quantum Interference Device. The electronic transport properties measurements were conducted in a Quantum Design Physical Properties Measurement System. Ohmic contacts were formed by using Al wire bonding connected to the sample surface.

***XRD and STEM Imaging:*** High resolution X-ray diffraction (HR-XRD) (Bruker D8, λ = 1.5406 Å) was used to characterize the crystallographic structures of thin films. STEM imaging was performed by a JEM-ARM200F (JEOL) microscope equipped with an ASCOR aberration corrector, a cold-field emission gun and a Gatan Quantum ER spectrometer, operated at 200 kV.

*X-ray diffraction of SFRO thin films*

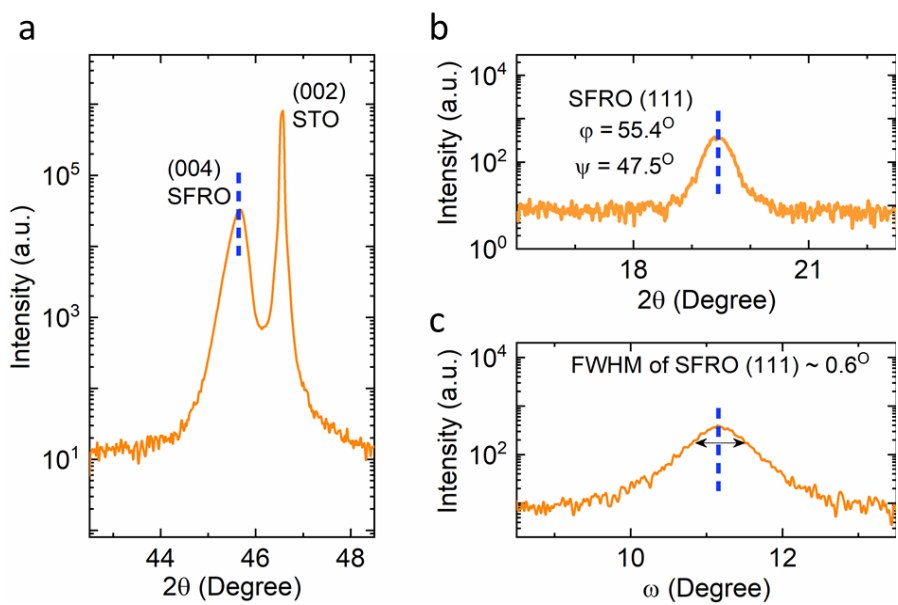

**Figure S1.** a. XRD θ-2θ scans for $Sr_2FeReO_6$ film on $SrTiO_3$ grown at 40 mTorr. b and c. Off-axis XRD θ-2θ scans near the $Sr_2FeReO_6$ 111 peak.

The ideal structure was defined in a tetragonal unit cell, space group I4/m, Z = 2, approximated by the ideal cubic structure of a double perovskite with the lattice parameter a = 7.864 Å, c = 7.901 Å. Figure S1a presents the refinement of the X-ray diffraction spectrum of the film grown under 40 mTorr at STO substrate. The spectra at room temperature indicate that the sample is single phase without the presence of tiny amounts of impurity phase. The peaks for SFRO were in agreement with a double perovskite structure, which indicates high-quality epitaxial films were prepared. It is known that the crystal structure of double perovskite SFRO consists of a regular arrangement of corner-sharing $FeO_6$ and $ReO_6$ octahedra alternating along the three directions of the crystal, with the large Sr cations occupying the voids between the octahedra. The order degree of $FeO_6$ and $ReO_6$ octahedra is also an important structural parameter. This ordering is verified by XRD off-axis scans at $\psi$ = 47.5° near the $Sr_2FeReO_6$ 111 peak. Its appearance confirms the Fe/Re ordering as shown in Figure S1b.

*Magnetoresistance of SFRO thin films*

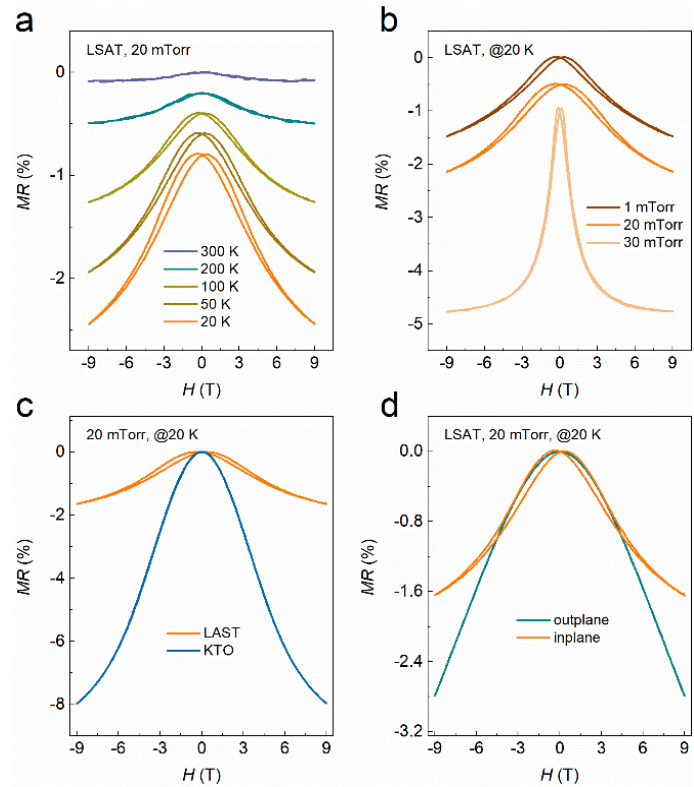

**Figure S2.** a. In-plane magnetoresistance of the SFRO film on LSAT obtained at 20 mTorr, measured at different temperatures with in-plane magnetic field. b. In-plane MR of SFRO films grown at various oxygen pressure at 20 K. c. In-plane MR of SFRO films grown at 20 mTorr on LSAT and KTO substrates at 20 K. d. MR of SFRO films grown at 20 mTorr on LSAT substrate with in-plane and out-of-plane magnetic field.

Magnetoresistance (*MR*) usually also gives meaningful magnetic properties. The filed dependent *MR* at different temperatures for LSAT-based films grown at 20 mTorr are presented in Figure S2a. Regardless of high or low temperature, a negative *MR* is observed. It is known that magnetotransport is dominated by spin-dependent carriers scattering at magnetic domain boundaries.[1,2] Since the hopping of spin-polarized electrons between magnetic domains is greatly affected by their relative magnetization directions, tunneling *MR* occurs at the boundaries, which may be controlled by an external magnetic field through the process of magnetic domain rotation. It is noted in Figure S2a that the value of *MR* of SFRO films increases with increasing magnetic field. The observed *MR* at 9 T, about 1.7 % at 20 K and 0.14 % at 300 K, are far less than those for polycrystalline samples and in coincidence with the low values of *MR* reported for single crystal films.[3,4] This might imply less carriers scattering at magnetic domain boundaries for our films than for polycrystalline case as an ideal crystal is expected to show no *MR* at the temperatures far below Curie temperature.[5] *MR* increases slowly with the decrease of temperature at 9 T, which suggests that grain boundary and magnetic domains mainly controls the *MR*. The *MR* values at low temperature are larger than those at high temperature due to the weak spin thermal fluctuation at low temperatures. Field dependent *MR* data at various $P_{O_2}$ are shown in Figure S2b. With increasing magnetic field, all the values of *MR* first increase sharply, then increase linearly in a slower trend. As been reported, there are two mechanisms which account for the *MR* of double perovskite.[6-7] One is the lowering of electron spin dependent crossing of magnetic domain boundaries under magnetic field. This mechanism is responsible for *MR* at low magnetic field, where the polarization of carriers is large. The other *MR* mechanism results from the quenching of spin scattering of the carriers by localized spins and prevails up to high magnetic field. The initial increase of *MR* at low field might be ascribed to first mechanism. The crossovers between the dramatic, low-field, and the gentle, high-field, *MR* processes coincide with the onset of the magnetization plateaus in the *M*(H) curves of Figure 2. We note that a linear correlation exists between the degree of cation ordering and the *MR* slope per unit magnetic field, assuming a linear variation of the *MR* with the applied magnetic field in the low field region and in the high field regime, as reported elsewhere.[2] Obviously, the 30 mTorr sample has a largest *MR* slope. Therefore, this result suggests a less antisite disorder in 30 mTorr sample, consistent with previous results.

Figure S2c shows field dependent *MR* data at different substrates. There is a butterfly curve with a reversible high-field slope except for the film on KTO substrate. This behavior for STO-based sample is contradictory to its weak magnetism. And it also contradicts a neat increase of the low-field *MR* per unit of applied magnetic field with saturated magnetic moment.[7] As we know, it is also apparent that there are other origins of an increased *MR* for double perovskite films, such as grain size and degraded surfaces. For our samples, the latter should be excluded because of similar synthesis conditions. It has been confirmed that STO based samples with low $P_{O_2}$ ($\leq$ 20 mTorr) prefer to be cation-disordered. This may have contributed to the prevalence of small grains or magnetic domain. The large *MR* for KTO based sample also supports this inference, consistent with polycrystalline samples. Typical in-plane *MR* hysteresis loops for LSAT based sample grown at 20 mTorr are shown in Figure S2d. The *MR* at 20 K shows distinct shape anisotropy. The difference between the hysteresis loops measured along the easy and hard axes is obvious, suggesting that the samples are provided with an in-plane uniaxial anisotropy. The surprising magnetic anisotropy is attributed to intrinsic magnetic anisotropy of the $Re^{5+}(5d^2)$ ions since little anisotropy is to be expected from the $Fe^{3+}$ ions with a half-filled d shell and no orbital moment in octahedral coordination. And shape anisotropy of films is the other possible reason. By making the easy axis lying in-plane, the demagnetization energy can be reduced, because the demagnetization field when magnetization is lying in-plane is smaller than when magnetization is lying out-of-plane.

## Temperature dependent magnetization of SFRO thin films

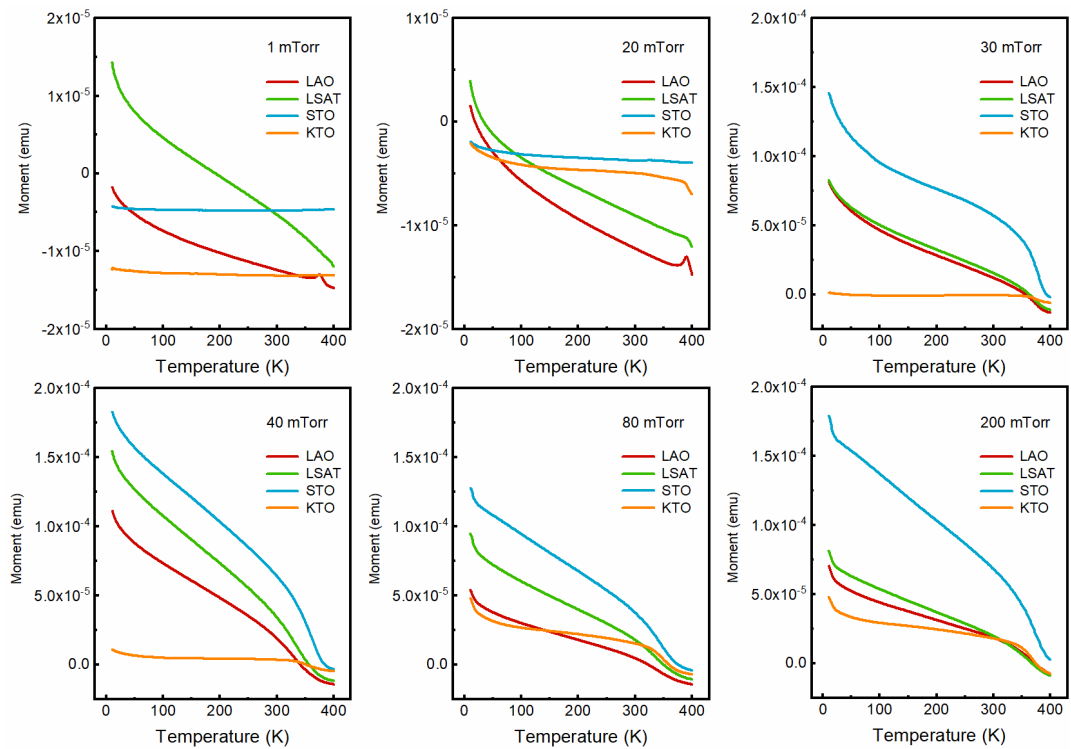

**Figure S3.** $M(T)$ curves with field cooling and measuring at 2000 Oe for $Sr_2FeReO_6$ films with different growth conditions.

## Normalized resistance of SFRO thin films

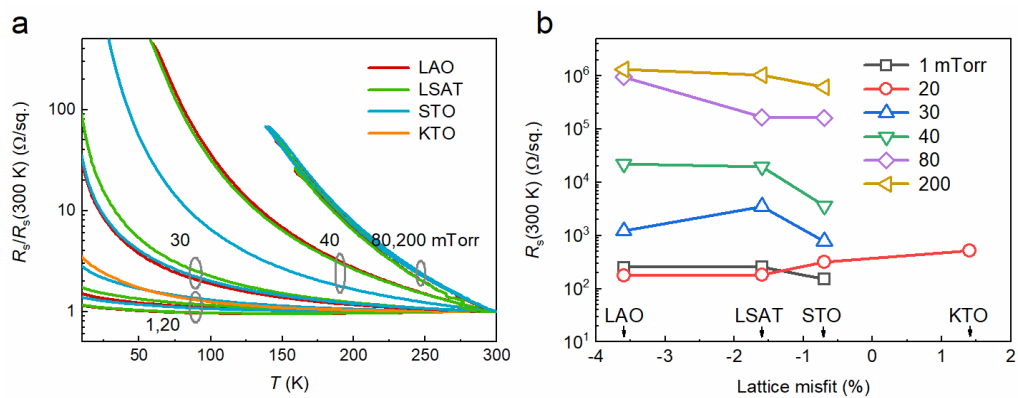

**Figure S4.** a and b. $R(T)$ curves and room-temperature resistance for $Sr_2FeReO_6$ films with different growth conditions.

S7